\documentclass[12pt,english]{article}
\usepackage[T1]{fontenc}
\usepackage[latin9]{inputenc}
\usepackage{geometry}
\usepackage{float}
\usepackage{amsmath}
\usepackage{amsthm}
\usepackage{amssymb}
\usepackage{graphicx}
\usepackage{setspace}
\usepackage[authoryear]{natbib}
\makeatletter

\providecommand{\tabularnewline}{\\}

\theoremstyle{definition}
\newtheorem{defn}{\protect\definitionname}
\theoremstyle{plain}
\newtheorem{prop}{\protect\propositionname}

\usepackage{tikz}
\usepackage{pgfplots}
\usepackage{xfrac}
\usepgfplotslibrary{fillbetween}
\setcitestyle{aysep={}}

\makeatother

\usepackage{babel}
\providecommand{\definitionname}{Definition}
\providecommand{\propositionname}{Proposition}

\begin{document}
\title{Comparing Electoral Polarization Levels\thanks{I thank Philipp Denter for helpful comments. Financial support from MICIN/AEI/10.13039/501100011033 grants CEX2021-001181-M, PID2020-118022GB-I00, and RYC2021-032163-I; and Comunidad de Madrid grants EPUC3M11 (V PRICIT) and H2019/HUM-5891 is gratefully acknowledged.}}
\author{Boris Ginzburg\thanks{Department of Economics, Universidad Carlos III de Madrid, Calle Madrid 126, 28903 Getafe (Madrid), Spain. Email: bginzbur@eco.uc3m.es.}}
\maketitle
\begin{abstract}
This paper introduces a definition of ideological polarization of an electorate around a particular central point. The definition is flexible about the location or boundaries of the center. Using US survey data, the paper shows how this approach can be used to establish whether polarization is occurring, and to find the position around which it is happening. I then show how ideological polarization as defined here is related to other phenomena, such as affective polarization and increased salience of divisive issues.

Keywords: polarization, ideology.
\end{abstract}
\newpage

Mass polarization is often described as ``disappearance of the center'' \citep{Abramowitz2010center}, or ``movement from the center toward the extremes'' \citep{fiorina2008polarization,levendusky2011red}. Given the significant consequences that polarization entails for political systems, it is not surprising that it has become a focus of much research. Yet detecting such a shift in policy preferences is difficult, because there is no single way of comparing distributions of voters' political positions. For example, much of the recent debate over whether the US electorate is becoming more ideologically polarized (see \citealt{fiorina2008polarization,abramowitz2008polarization,westfall2015perceiving,menchaca2023americans}) is due to the different definitions of polarization \citep{lelkes2016mass,mehlhaff2022group}. 


When voters have clear group identities, such as party affiliations, polarization may be evaluated by measuring ideological distance between groups and the degree of homogeneity within them \citep{esteban1994measurement,duclos2004polarization,mehlhaff2022group}, or the degree of overlap between group-specific distributions of voters' positions \citep{levendusky2011red,lelkes2016mass}. However, such indicators are of little help when voters' group identities are not clearly defined, or when one needs to measure ideological polarization across a heterogeneous electorate, rather than between specific groups.


To measure mass polarization without relying on exogenous group identities, one needs a measure of dispersion around a certain central point. For this, many papers focus on statistical moments of the distribution of voters' positions. These include variance or standard deviation \citep{dimaggio1996have,ezrow2007variance,levendusky2009microfoundations}; kurtosis \citep{dimaggio1996have,baldassarri2007dynamics}; or bimodality coefficient, which is a function of skewness and kurtosis \citep{lelkes2016mass}. By construction, all these indicators measure polarization as dispersion around the mean. For some applications, it is reasonable to assume that the mean is the relevant central point. But for others, it is not. For example, in formal models that follow \citet{Downs_Anthony1957}, it often makes more sense to think of dispersion around the median. In models of information manipulation (see \citealp{rosenfeld2024information} for an overview), the central point might be the position of a voter who is undecided in the absence of further information. When working with survey data, a researcher may think that a centrist is a voter who gives a ``neutral'' response -- for example, who neither likes nor dislikes a certain policy proposal or a politician. In all of these cases, the centrist position is not, in general, the average position.

Other studies use survey data to measure polarization through a change in the share of voters that identify as centrists. \citet{fiorina2008political} and \citet{iversen2015information} consider as centrists those voters who placed themselves exactly in the middle of, respectively, a 7-point and an 11-point left-right scale. \citet{levendusky2009microfoundations} focuses on those who place themselves in a wider 3-5 interval on a 7-point scale. \citet{abramowitz2008polarization} label as moderates those with a score of 2 or 3 on a 0-7 aggregate scale. Thus, the definition of centrists can vary, and such measures require the researcher to define how far the left and the right boundaries of the centrist interval are. Consequently, the location of these boundaries can have a considerable effect on the results of the analysis.

In this paper, I propose a way of evaluating polarization that operationalizes the idea of the disappearance of the center. Crucially, the model allows the researcher to select the centrist position depending on a specific application, and does not make assumptions about the width of the centrist interval. As a starting point, I take an electorate in which each voter has an ideal policy on a unidimensional policy space. A single point $x^{*}$ -- which may be the mean, the median, or any other position relevant to the context -- is taken as the centrist position. I then introduce a definition of\emph{ polarization around $x^{*}$}. Specifically, I define an electorate to be more polarized around $x^{*}$ if the share of voters belonging to \emph{any} interval that includes $x^{*}$ is smaller. Thus, polarization increases if the fraction of centrist voters falls, \emph{regardless of how restrictively one defines the centrists}. The paper then proves a simple necessary and sufficient condition that can be easily applied to a pair of distributions to determine whether one represents greater polarization around a particular position $x^{*}$.

Using US survey data as an example, I then show how this model can help establish whether polarization around a given $x^{*}$ has been rising. Furthermore, I show how, by being agnostic about the central position $x^{*}$, this measure of polarization, unlike existing measures, can be used to determine \emph{which} position the voters are becoming more polarized around.

Furthermore, I show how this definition is related to several features of polarization. First, it helps link ideological polarization described above to affective polarization -- increased dislike towards members of the opposing political group. Past research \citep{rogowski2016ideology,bougher2017correlates,webster2017ideological,orr2020policy} suggests that increased ideological divide may drive affective polarization. This paper provides a formal model consistent with these empirical results. I model a setting in which $x^{*}$ serves as a boundary between two political groups. Each voter's animosity towards the opposite group is an increasing function of ideological distance between her and the average member of that group. The paper shows that increased ideological polarization around $x^{*}$ implies increased average level of animosity.

At the same time, there is some evidence \citep{han2023issue} that increased salience of divisive issues may be related to increase in affective polarization. The framework for analyzing polarization developed here provides a mechanism consistent with this observation. I consider a formal model in which a voter's political position on the left-right spectrum is a weighted average of her positions on two issues. The paper shows that an increase in the weight of the more divisive issue implies increased ideological polarization as defined in this paper, and hence also increased affective polarization.

\section*{A Model of Polarization }

Consider a continuum of voters, and an interval $X\subset\mathbb{R}$ of policies. Let $F$ be the cumulative distribution of voters' ideal policies over $X$. To fix ideas, we will say that the higher the value of $x$, the more right-wing the voter is. 

Let $x^{*}\in X$ be the policy that a researcher considers centrist. One might think of centrist voters as those whose ideal policies are close to $x^{*}$, that is, fall into some interval $\left[\underline{x},\overline{x}\right]$ that includes $x^{*}$. Polarization then involves a reduction in the fraction of voters whose policies belong to the centrist interval $\left[\underline{x},\overline{x}\right]$, and an increase in the fraction of voters whose policies lie outside of it. However, such a definition relies on an arbitrary choice of the boundaries $\underline{x}$ and $\overline{x}$. Instead, we can define polarization as a reduction in the fraction of voters who are centrists, \emph{whichever way the boundaries of the center are set:}
\begin{defn}
\label{def:polarisation}Consider two distributions $F$ and $\hat{F}$, and a policy $x^{*}\in X$. We say that polarization around \emph{$x^{*}$ }is higher\emph{ }under $\hat{F}$ than under $F$ if and only if for any interval $\left[\underline{x},\overline{x}\right]$ that includes $x^{*}$, the share of voters whose ideal policies belong to $\left[\underline{x},\overline{x}\right]$ is weakly smaller under $\hat{F}$ than under $F$.
\end{defn}
Thus, increased polarization means a reduction in the fraction of centrists, and an increase in the fraction of radicals, wherever the dividing lines between the former and the latter are set. Note that the definition is specific to a given central position: it describes polarization around a particular point $x^{*}$. For example, polarization may increase around the median position. Or instead there may be an increase in polarization around a moderate right position, that is, between the extreme right and other voters. As will be shown later, this flexibility is helpful for determining the point, if any, around which polarization occurs.



A downside of this definition is that it is difficult to work with -- by looking at a pair of distributions, it is not easy to say whether one admits higher polarization than the other. However, the following result (the proof of which, like those of other propositions, is in the Appendix) provides a simple necessary and sufficient condition to establish this:
\begin{prop}
\label{prop: centre}Polarization around \emph{$x^{*}$ }is higher under distribution $\hat{F}$ than under distribution $F$ if and only if both of the following statements are true: (i) $\hat{F}\left(x\right)-F\left(x\right)\geq0$ for all $x\leq x^{*}$, and (ii) $\hat{F}\left(x\right)-F\left(x\right)\leq0$ for all $x\geq x^{*}$.
\end{prop}
Hence, $\hat{F}$ implies higher polarization around $x^{*}$ than $F$ if $\hat{F}\left(x\right)-F\left(x\right)$ crosses zero exactly once, at $x^{*}$. Equivalently, polarization around $x^{*}$ is higher under $\hat{F}$ than under $F$ if $F$ first-order stochastically dominates $\hat{F}$ for $x<x^{*}$, while the reverse holds for $x>x^{*}$.

\section*{Example: Polarization of the US Electorate}

To see how this definition can be used to analyze polarization, we can take the data from the American National Election Studies database. Table \ref{tab:shares} in the  Appendix presents the distribution of respondents' answers when asked to place themselves on a scale from 0 to 10, where 0 means the left and 10 means the right, for the years 1996, 2004, and 2016.



\begin{figure}
\centering{}%
\begin{minipage}[t]{0.47\columnwidth}%
\begin{center}
\includegraphics[scale=0.15]{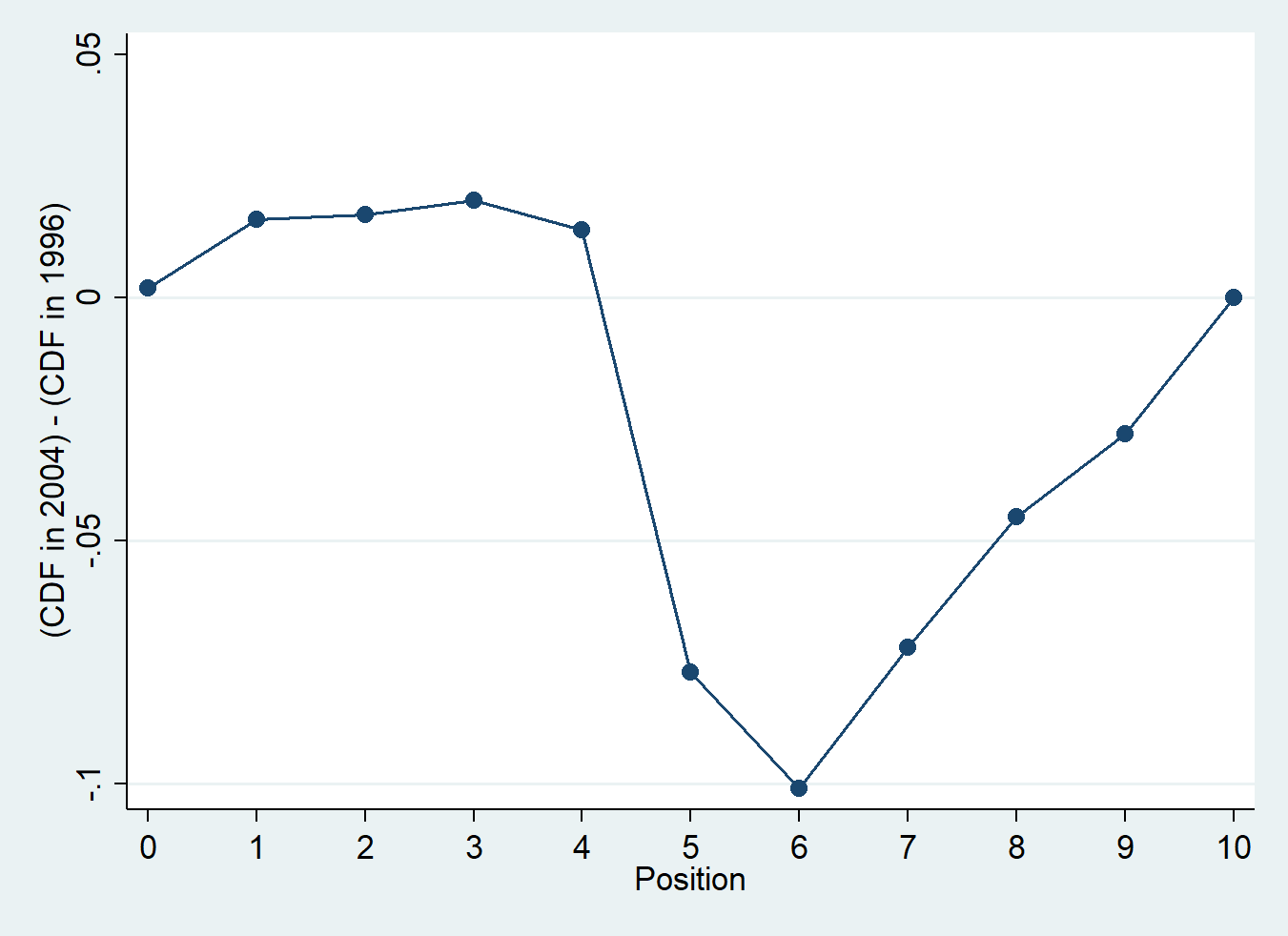}
\par\end{center}%
\end{minipage}%
\begin{minipage}[t]{0.47\columnwidth}%
\begin{center}
\includegraphics[scale=0.15]{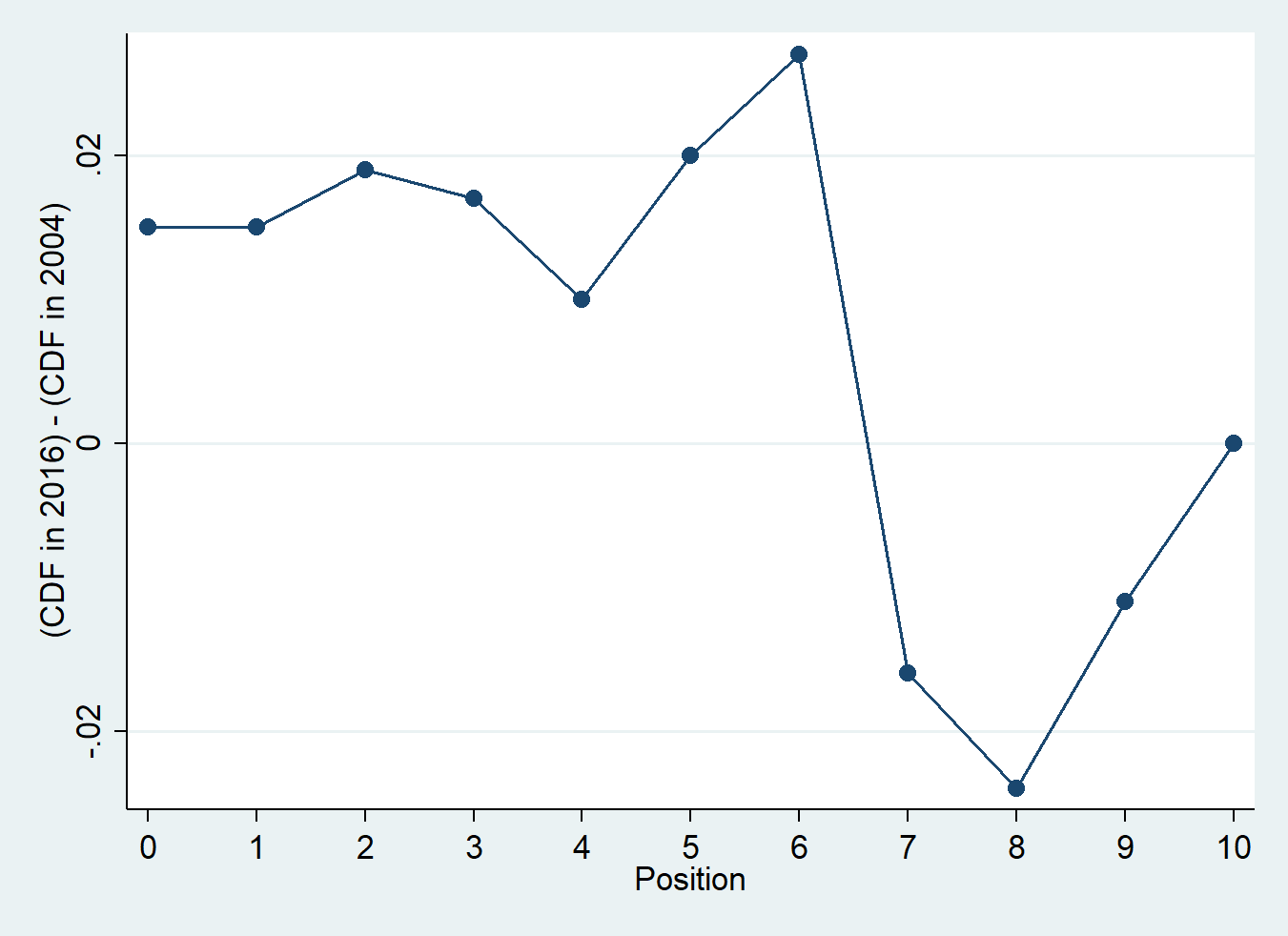}
\par\end{center}%
\end{minipage}\caption{\label{fig:CDFs}Left: difference between the CDF of voters' positions in 2004 and in 1996. Right: difference between the CDF of voters' positions in 2016 and in 2004.}
\end{figure}

The left panel of Figure \ref{fig:CDFs} shows the difference between the cumulative distribution functions of political positions in 2004 and in 1996. It is positive to the left of a point between 4 and 5, and negative to the right of that point. By Proposition \ref{prop: centre}, this implies that the electorate became more polarized around some $x^{*}$ close to 5. This is consistent with conventional measures of polarization: Table \ref{tab:shares} shows that the share of respondents who place themselves in the middle of the scale has decreased, while the variance of responses has increased. 


Now consider how the distribution has evolved between 2004 and 2016. Here conventional measures of polarization do not yield a clear answer. On the one hand, the number of respondents placing themselves in the exact middle, at 5, has increased. The proportion of respondents between 4 and 6 was also somewhat greater in 2016 than in 2004 (respectively, 46.48\% and 45.46\%). This suggests a moderate decrease in polarization. On the other hand, variance has risen from 2004 to 2016, suggesting increased polarization.

What do Definition \ref{def:polarisation} and Proposition \ref{prop: centre} say? The right panel of Figure \ref{fig:CDFs} displays the difference between the CDFs in 2016 and in 2004. The difference does not cross zero close to 5. It does, however, cross zero once, between 6 and 7, indicating increased polarization around that point. Thus, while we cannot conclude that polarization between left-leaning and right-leaning voters rose or fell, we do observe an increase in polarization between right-wingers (those who place themselves to the right of 6) and everyone else.

Thus, by being flexible about the location of the central point $x^{*}$, the framework introduced here can determine the point, if any, around which polarization increases or decreases -- unlike conventional measures of polarization, which assume a particular central point. 


\part*{Discussion\label{sec:Discussion}}

\paragraph{Ideological polarization and affective polarization.}

A frequent topic of research is affective polarization: the dislike towards members of the opposite political group.\footnote{See \citet{iyengar2019origins} for an overview.} How does ideological polarization defined above relate to affective polarization?

Let $x^{*}$ be a dividing line between two political groups, the left ($L$) and the right ($R$). Voters with $x<x^{*}$ belong to $L$, and voters with $x>x^{*}$ belong to $R$. Suppose that a voter forms a more negative opinion about the opposing group if that group is on average more dissimilar to her. Formally, suppose that for each voter with ideal policy $x\in X$ belonging to group $i\in\left\{ L,R\right\} $, the animosity towards the opposing group $j\neq i$ equals
\[
g\left(\left|x-m_{j}\right|\right),
\]
where $g\left(\cdot\right)$ is some increasing function, $\left|\cdot\right|$ is the absolute value, and $m_{j}$ is the position of the average member of group $j$. Let the overall level of affective polarization be the average value of animosity across all voters. Then we can show the following result:
\begin{prop}
\label{prop: affective}Take any increasing function $g\left(\cdot\right)$. An increase in polarization around $x^{*}$ according to Definition \ref{def:polarisation} implies an increase in affective polarization.
\end{prop}
In words, greater ideological polarization  around the boundary between the two political groups implies a higher level of affective polarization between these groups. 


\paragraph{Polarization and salience of divisive issues.}

Some issues are more divisive than others (see, for instance, \citealp{ash2017elections}). For example, there is evidence that the US electorate is more divided on social issues than on economic issues \citep{saad2021americans}. Suppose that the position $x\in X$ of a voter is determined by her stance on two issues, one of which is ``divisive'', while the other is ``common-value''. Specifically, suppose that for each voter, we have 
\[
x=\left(1-\alpha\right)c+\alpha d,
\]
where $c$ and $d$ are real numbers, and $\alpha\in\left[0,1\right]$. The numbers $c$ and $d$ are drawn independently from distributions $G_{c}$ and $G_{d}$, respectively. The number $c$ represents the voter's stance on the common-value issue, while $d$ represents her stance of the divisive issue. The two issues differ in the following way: $d$ can take any value on $X$, while $c$ can only take values on some interval $\left[\underline{c},\overline{c}\right]\subset X$. In words, the range of voters' positions on the common-value issue is more narrow than the range of positions on the divisive issue. The coefficient $\alpha$ measures the salience of the divisive issue.\footnote{In the limiting case $\underline{c}$ and $\overline{c}$ are arbitrarily close to each other and equal the expected value of $d$, a decrease in $\alpha$ is equivalent to a ``squeeze'' of the distribution of $x$ in the language of \citet{duclos2004polarization}.} We can then show that when $\underline{c}$ and $\overline{c}$ are close -- that is, when voters' views on the common-value issue tend to be similar -- increased salience of the divisive issue makes the electorate more polarized over the overall ideological distribution $F$:
\begin{prop}
\label{prop: salience}If $\underline{c}$ and $\overline{c}$ are sufficiently close to each other, an increase in $\alpha$ increases polarization of $x$ around some $x^{*}\in\left[\underline{c},\overline{c}\right]$.
\end{prop}
Recall that by Proposition \ref{prop: affective}, increased ideological polarization implies increased affective polarization. Hence, greater salience of divisive issues increases affective polarization.



\paragraph{A continuous measure of polarization.}

One limitation of the approach introduced in this paper is that it creates only a partial order of distributions. If the difference between two distributions crosses zero multiple times, they cannot be compared in terms of polarization. 

An intuitive way of extending the definition is to say that a shift of the distribution of ideal policies from $F$ to $\hat{F}$ increases polarization around $x^{*}$ whenever the condition $\hat{F}\left(x\right)-F\left(x\right)\geq0$ ``tends to be hold'' for $x\leq x^{*}$, and ``tends to fail'' for $x\geq x^{*}$. Thus, increased polarization means that the distribution ``tends to have'' higher values for $x\leq x^{*}$, and lower values for $x\geq x^{*}$. Formally, we can measure the degree of polarization of distribution $F$ around $x^{*}$ as
\[
P\left(F,x^{*}\right):=a\left[\int_{x<x^{*}}F\left(x\right)dx\right]-b\left[\int_{x>x^{*}}F\left(x\right)dx\right],
\]
where $a\left[\cdot\right]$ and $b\left[\cdot\right]$ are strictly increasing functions that can potentially depend on $x^{*}$ and $F\left(x^{*}\right)$ (but not on $x$ or on the shape of $F$ more generally). Note that if $F$ increases for $x<x^{*}$, or if $F$ decreases for $x>x^{*}$, the function $P\left(F,x^{*}\right)$ increases. Hence, increased polarization around $x^{*}$ according to the earlier definition implies an increase in $P\left(F,x^{*}\right)$. At the same time, $P\left(F,x^{*}\right)$ nests as special cases a number of intuitive measures of polarization. For example, when $x^{*}$ is a boundary between the Left and the Right as proposed above, then the difference between between the average position of a left-wing voter and that of a right-wing voter is a special case of $P\left(F,x^{*}\right)$.\footnote{Specifically, this is the case when $a\left[y\right]=\frac{y-x^{*}F\left(x^{*}\right)}{F\left(x^{*}\right)}$ and $b\left[y\right]=\frac{y-\left[1-x^{*}F\left(x^{*}\right)\right]}{1-F\left(x^{*}\right)}$.}

\section*{Conclusions}

Ideological polarization is a topic of considerable research, which, however, is complicated by a lack of consensus on how to define polarization. This paper has proposed a way of comparing distributions of voters' ideological positions in terms of polarization around a given point. This approach allows the researcher considerable flexibility in determining the point around which polarization occurs. At the same time, it provides a framework for linking ideological polarization, affective polarization, and salience of particular policy issues.

One limitation of this approach is that it creates only a partial order of distributions. While the paper has suggested a way of constructing a continuous measure of polarization, future research can consider which functions $a\left[\cdot\right]$ and $b\left[\cdot\right]$ are best suited for specific applications.

\appendix
\renewcommand{\thetable}{A\arabic{table}} 

\section*{Appendix}

\subsection*{Additional Tables}

\begin{table}[H]
\centering{}%
\begin{tabular}{|c|ccc|}
\hline 
\textbf{Political position} & \textbf{Share in 1996} & \textbf{Share in 2004} & \textbf{Share in 2016}\tabularnewline
\hline 
\textbf{0 (Left)} & 1.002\% & 1.169\% & 2.734\%\tabularnewline
\textbf{1} & 1.336\% & 2.682\% & 2.623\%\tabularnewline
\textbf{2} & 4.285\% & 4.478\% & 4.845\%\tabularnewline
\textbf{3} & 5.920\% & 6.211\% & 6.007\%\tabularnewline
\textbf{4} & 7.911\% & 7.316\% & 6.717\%\tabularnewline
\textbf{5} & 37.365\% & 28.280\% & 29.167\%\tabularnewline
\textbf{6} & 12.236\% & 9.868\% & 10.594\%\tabularnewline
\textbf{7} & 11.231\% & 14.086\% & 9.796\%\tabularnewline
\textbf{8} & 9.756\% & 12.395\% & 11.612\%\tabularnewline
\textbf{9} & 3.751\% & 5.558\% & 6.815\%\tabularnewline
\textbf{10 (Right)} & 5.207\% & 7.956\% & 9.090\%\tabularnewline
\hline 
\textbf{Average position} & 5.620 & 5.875 & 5.803\tabularnewline
\textbf{Variance of positions} & 4.133 & 5.336 & 6.108\tabularnewline
\textbf{Sample} & 1408 & 917 & 3502\tabularnewline
\hline 
\end{tabular}\caption{\label{tab:shares}Shares of respondents by self-reported position on a 0-10 left-right axis, by year. Source: ANES dataset, question VCF9240. Answers are weighted using the weights provided by ANES. Respondents who did not know how to answer or refused to answer are omitted.}
\end{table}

\subsection*{Proofs}

\paragraph{Proof of Proposition \ref{prop: centre}.}

For any interval $\left[\underline{x},\overline{x}\right]$, the fraction of voters whose ideal policies belong to that interval equals $F\left(\overline{x}\right)-F\left(\underline{x}\right)$. Hence, polarization around $x^{*}$ is higher under $\hat{F}$ than under $F$ if and only if
\begin{equation}
\hat{F}\left(\overline{x}\right)-\hat{F}\left(\underline{x}\right)\leq F\left(\overline{x}\right)-F\left(\underline{x}\right)\text{ for all }\underline{x},\overline{x}\text{ such that }\underline{x}\leq x^{*}\leq\overline{x}\label{eq:centre reduction}
\end{equation}
To prove the proposition, we need to prove that for any $x^{*}\in X$, the two conditions given there hold if and only if (\ref{eq:centre reduction}) holds.

Take some $x^{*}\in X$. To prove one direction of the statement, suppose (\ref{eq:centre reduction}) holds. When $\underline{x}=\min\left\{ X\right\} $, we have $\hat{F}\left(\underline{x}\right)=F\left(\underline{x}\right)=0$, which together with (\ref{eq:centre reduction}) implies that $\hat{F}\left(\overline{x}\right)\leq F\left(\overline{x}\right)$ for every $\overline{x}\geq x^{*}$. When $\overline{x}=\max\left\{ X\right\} $, we have $\hat{F}\left(\overline{x}\right)=F\left(\overline{x}\right)=1$, which together with (\ref{eq:centre reduction}) implies that $\hat{F}\left(\underline{x}\right)\geq F\left(\underline{x}\right)$ for every $\underline{x}\leq x^{*}$.

To prove the other direction, suppose that $\hat{F}\left(\underline{x}\right)\geq F\left(\underline{x}\right)$ for all $\underline{x}\leq x^{*}$, and $\hat{F}\left(\overline{x}\right)\leq F\left(\overline{x}\right)$ for all $\overline{x}\geq x^{*}$. Subtracting the first inequality from the second, we obtain (\ref{eq:centre reduction}).\qed

\paragraph{Proof of Proposition \ref{prop: affective}.}

Suppose polarization around $x^{*}$ is higher under $F$ than under $\hat{F}$ according to Definition \ref{def:polarisation}. Proposition \ref{prop: centre} implies that $F$ first-order stochastically dominates $\hat{F}$ for $x<x^{*}$, and $\hat{F}$ first-order stochastically dominates $F$ for $x>x^{*}$. For each group $j\in\left\{ L,R\right\} $, let $m_{j}$ and $\hat{m}_{j}$ be the average ideal policies of members of the group under distributions $F$ and $\hat{F}$, respectively. By a well-known property of stochastic dominance, we have $m_{L}\geq\hat{m}_{L}$ and $\hat{m}_{R}\geq m_{R}$. 



Let $A\left(F\right)$ and $A\left(\hat{F}\right)$ be the levels of affective polarization under $F$ and under $\hat{F}$, respectively. We have

\begin{align*}
A\left(F\right)= & \int_{x<x^{*}}g\left(m_{R}-x\right)dF\left(x\right)+\int_{x>x^{*}}g\left(x-m_{L}\right)dF\left(x\right)\\
\leq & \int_{x<x^{*}}g\left(m_{R}-x\right)d\hat{F}\left(x\right)+\int_{x>x^{*}}g\left(x-m_{L}\right)d\hat{F}\left(x\right)\\
\leq & \int_{x<x^{*}}g\left(\hat{m}_{R}-x\right)d\hat{F}\left(x\right)+\int_{x>x^{*}}g\left(x-\hat{m}_{L}\right)d\hat{F}\left(x\right)=A\left(\hat{F}\right).
\end{align*}
The first inequality holds due to the stochastic dominance relation between $\hat{F}$ and $F$, and the fact that $g\left(m_{R}-x\right)$ is decreasing in $x$ while $g\left(x-m_{L}\right)$ is increasing in $x$. The second inequality follows from the fact that $m_{L}\geq\hat{m}_{L}$ and $\hat{m}_{R}\geq m_{R}$. Thus, $A\left(\hat{F}\right)\geq A\left(F\right)$.\qed


\paragraph{Proof of Proposition \ref{prop: salience}.}

Since for each voter, $x=\left(1-\alpha\right)c+\alpha d$, the distribution of $x$ is given by 
\[
F\left(x\right)=\Pr\left[\left(1-\alpha\right)c+\alpha d<x\right]=\Pr\left[d<\frac{x-\left(1-\alpha\right)c}{\alpha}\right]=\int_{\underline{c}}^{\overline{c}}G_{d}\left[\frac{x-\left(1-\alpha\right)c}{\alpha}\right]dG_{c}\left(c\right).
\]
For a given $x\in X$, an increase in $\alpha$ increases $F\left(x\right)$ if and only if $\frac{\partial F\left(x\right)}{\partial\alpha}>0$, that is, if and only if
\begin{equation}
\int_{\underline{c}}^{\overline{c}}g_{d}\left[\frac{x-\left(1-\alpha\right)c}{\alpha}\right]\frac{c-x}{\alpha^{2}}dG_{c}\left(c\right)>0,\label{eq:derivative wrt alpha}
\end{equation}
where $g_{d}$ is the density of $d$.

When $x<\underline{c}$, we have $g_{d}\left[\frac{x-\left(1-\alpha\right)c}{\alpha}\right]\frac{c-x}{\alpha^{2}}dG_{c}\left(c\right)>0$ for all $c\in\left[\underline{c},\overline{c}\right]$, so (\ref{eq:derivative wrt alpha}) holds. When $x>\overline{c}$, we have $g_{d}\left[\frac{x-\left(1-\alpha\right)c}{\alpha}\right]\frac{c-x}{\alpha^{2}}dG_{c}\left(c\right)<0$ for all $c\in\left[\underline{c},\overline{c}\right]$, so (\ref{eq:derivative wrt alpha}) does not hold. Now take $x\in\left[\underline{c},\overline{c}\right]$. For these values of $x$, the left-hand side of (\ref{eq:derivative wrt alpha}) is decreasing in $x$ if and only if its derivative with respect to $x$ is negative, that is, if and only if
\begin{align}
 & \int_{\underline{c}}^{\overline{c}}g_{d}^{\prime}\left[\frac{x-\left(1-\alpha\right)c}{\alpha}\right]\frac{c-x}{\alpha^{3}}dG_{c}\left(c\right)-\int_{\underline{c}}^{\overline{c}}g_{d}\left[\frac{x-\left(1-\alpha\right)c}{\alpha}\right]\frac{1}{\alpha^{2}}dG_{c}\left(c\right)<0\nonumber \\
\iff & E\left(g_{d}^{\prime}\left[\frac{x-\left(1-\alpha\right)c}{\alpha}\right]\frac{c-x}{\alpha}\right)<E\left(g_{d}\left[\frac{x-\left(1-\alpha\right)c}{\alpha}\right]\right),\label{eq:condition for salience}
\end{align}
where $E$ denotes the expectation over $G_{c}$. If $\underline{c}$ and $\overline{c}$ are sufficiently close to each other, then $c-x$ is arbitrarily small for all $x\in\left[\underline{c},\overline{c}\right]$. In this case, the left-hand side of (\ref{eq:condition for salience}) is arbitrarily close to zero, while the right-hand side is strictly positive. Hence, (\ref{eq:condition for salience}) holds. Thus, when $\underline{c}$ and $\overline{c}$ are sufficiently close to each other, the left-hand side of (\ref{eq:derivative wrt alpha}) is decreasing in $x$ for all $x\in\left[\underline{c},\overline{c}\right]$. Together with the fact that (\ref{eq:derivative wrt alpha}) holds for all $x<\underline{c}$ and does not hold for all $x>\overline{c}$, this means that there exists $x^{*}\in\left[\underline{c},\overline{c}\right]$ such that (\ref{eq:derivative wrt alpha}) holds if and only if $x<x^{*}$. Hence, an increase in $\alpha$ increases $F\left(x\right)$ for $x<x^{*}$, and decreases $F\left(x\right)$ for $x>x^{*}$, implying an increase in polarization around $x^{*}$.\qed

\begin{singlespace}
\bibliography{polarisation}

@Book{Downs_Anthony1957,
  author    = {Downs, Anthony},
  publisher = {Harper and Row},
  title     = {An Economic Theory of Democracy},
  year      = {1957},
  edition   = {Paperback},
  isbn      = {978-0060417505},
  date      = {1957},
  isbn10    = {0060417501},
  language  = {English},
  pages     = {310},
}

@Article{duclos2004polarization,
  author    = {Duclos, Jean-Yves and Esteban, Joan and Ray, Debraj},
  journal   = {Econometrica},
  title     = {Polarization: Concepts, Measurement, Estimation},
  year      = {2004},
  number    = {6},
  pages     = {1737--1772},
  volume    = {72},
  publisher = {The Econometric Society},
}

@Article{abramowitz2008polarization,
  author    = {Abramowitz, Alan I and Saunders, Kyle L},
  journal   = {The Journal of Politics},
  title     = {Is polarization a myth?},
  year      = {2008},
  number    = {2},
  pages     = {542--555},
  volume    = {70},
  publisher = {Cambridge University Press New York, USA},
}

@Article{bougher2017correlates,
  author    = {Bougher, Lori D},
  journal   = {Political Behavior},
  title     = {The correlates of discord: identity, issue alignment, and political hostility in polarized America},
  year      = {2017},
  number    = {3},
  pages     = {731--762},
  volume    = {39},
  publisher = {Springer},
}

@Article{rogowski2016ideology,
  author    = {Rogowski, Jon C and Sutherland, Joseph L},
  journal   = {Political Behavior},
  title     = {How ideology fuels affective polarization},
  year      = {2016},
  pages     = {485--508},
  volume    = {38},
  publisher = {Springer},
}

@Article{webster2017ideological,
  author    = {Webster, Steven W and Abramowitz, Alan I},
  journal   = {American Politics Research},
  title     = {The ideological foundations of affective polarization in the US electorate},
  year      = {2017},
  number    = {4},
  pages     = {621--647},
  volume    = {45},
  publisher = {SAGE Publications Sage CA: Los Angeles, CA},
}

@Article{iyengar2019origins,
  author    = {Iyengar, Shanto and Lelkes, Yphtach and Levendusky, Matthew and Malhotra, Neil and Westwood, Sean J},
  journal   = {Annual Review of Political Science},
  title     = {The origins and consequences of affective polarization in the United States},
  year      = {2019},
  pages     = {129--146},
  volume    = {22},
  publisher = {Annual Reviews},
}

@Article{fiorina2008polarization,
  author    = {Fiorina, Morris P and Abrams, Samuel A and Pope, Jeremy C},
  journal   = {The Journal of Politics},
  title     = {Polarization in the American public: Misconceptions and misreadings},
  year      = {2008},
  number    = {2},
  pages     = {556--560},
  volume    = {70},
  publisher = {Cambridge University Press New York, USA},
}

@Article{westfall2015perceiving,
  author    = {Westfall, Jacob and Van Boven, Leaf and Chambers, John R and Judd, Charles M},
  journal   = {Perspectives on Psychological Science},
  title     = {Perceiving political polarization in the United States: Party identity strength and attitude extremity exacerbate the perceived partisan divide},
  year      = {2015},
  number    = {2},
  pages     = {145--158},
  volume    = {10},
  publisher = {Sage Publications Sage CA: Los Angeles, CA},
}

@Article{mehlhaff2022group,
  author    = {Mehlhaff, Isaac D},
  journal   = {American Political Science Review},
  title     = {A Group-Based Approach to Measuring Polarization},
  year      = {2022},
  pages     = {1--9},
  publisher = {Cambridge University Press},
}

@Article{orr2020policy,
  author    = {Orr, Lilla V and Huber, Gregory A},
  journal   = {American Journal of Political Science},
  title     = {The policy basis of measured partisan animosity in the United States},
  year      = {2020},
  number    = {3},
  pages     = {569--586},
  volume    = {64},
  publisher = {Wiley Online Library},
}

@Article{han2023issue,
  author    = {Han, Kyung Joon},
  journal   = {Journal of Elections, Public Opinion and Parties},
  title     = {Issue salience and affective polarization},
  year      = {2023},
  pages     = {1--19},
  publisher = {Taylor \& Francis},
}

@Article{esteban1994measurement,
  author    = {Esteban, Joan-Maria and Ray, Debraj},
  journal   = {Econometrica: Journal of the Econometric Society},
  title     = {On the measurement of polarization},
  year      = {1994},
  pages     = {819--851},
  publisher = {JSTOR},
}

@Book{Abramowitz2010center,
  author    = {Alan I. Abramowitz},
  publisher = {Yale University Press},
  title     = {The Disappearing Center: Engaged Citizens, Polarization, and American Democracy},
  year      = {2010},
  address   = {New Haven},
}

@Article{lelkes2016mass,
  author    = {Lelkes, Yphtach},
  journal   = {Public Opinion Quarterly},
  title     = {Mass polarization: Manifestations and measurements},
  year      = {2016},
  number    = {S1},
  pages     = {392--410},
  volume    = {80},
  publisher = {Oxford University Press US},
}

@Article{menchaca2023americans,
  author    = {Menchaca, Marcos},
  journal   = {Journal of Elections, Public Opinion and Parties},
  title     = {Are Americans polarized on issue dimensions?},
  year      = {2023},
  number    = {2},
  pages     = {228--246},
  volume    = {33},
  publisher = {Taylor \& Francis},
}

@Article{dimaggio1996have,
  author    = {DiMaggio, Paul and Evans, John and Bryson, Bethany},
  journal   = {American Journal of Sociology},
  title     = {Have Americans' social attitudes become more polarized?},
  year      = {1996},
  number    = {3},
  pages     = {690--755},
  volume    = {102},
  publisher = {University of Chicago Press},
}

@Article{baldassarri2007dynamics,
  author    = {Baldassarri, Delia and Bearman, Peter},
  journal   = {American Sociological Review},
  title     = {Dynamics of political polarization},
  year      = {2007},
  number    = {5},
  pages     = {784--811},
  volume    = {72},
  publisher = {Sage Publications Sage CA: Los Angeles, CA},
}

@Article{ezrow2007variance,
  author    = {Ezrow, Lawrence},
  journal   = {The Journal of Politics},
  title     = {The variance matters: How party systems represent the preferences of voters},
  year      = {2007},
  number    = {1},
  pages     = {182--192},
  volume    = {69},
  publisher = {Cambridge University Press New York, USA},
}

@Article{iversen2015information,
  author    = {Iversen, Torben and Soskice, David},
  journal   = {Comparative Political Studies},
  title     = {Information, inequality, and mass polarization: Ideology in advanced democracies},
  year      = {2015},
  number    = {13},
  pages     = {1781--1813},
  volume    = {48},
  publisher = {SAGE Publications Sage CA: Los Angeles, CA},
}

@Article{levendusky2009microfoundations,
  author    = {Levendusky, Matthew S},
  journal   = {Political Analysis},
  title     = {The microfoundations of mass polarization},
  year      = {2009},
  number    = {2},
  pages     = {162--176},
  volume    = {17},
  publisher = {Cambridge University Press},
}

@Article{levendusky2011red,
  author    = {Levendusky, Matthew S and Pope, Jeremy C},
  journal   = {Public Opinion Quarterly},
  title     = {Red states vs. blue states: going beyond the mean},
  year      = {2011},
  number    = {2},
  pages     = {227--248},
  volume    = {75},
  publisher = {Oxford University Press},
}

@Article{fiorina2008political,
  author  = {Fiorina, Morris and Abrams, Samuel J},
  journal = {Annual Review of Political Science},
  title   = {Political Polarization in the American Public},
  year    = {2008},
  volume  = {11},
}

@Article{rosenfeld2024information,
  author    = {Rosenfeld, Bryn and Wallace, Jeremy},
  journal   = {Annual Review of Political Science},
  title     = {Information Politics and Propaganda in Authoritarian Societies},
  year      = {2024},
  volume    = {27},
  publisher = {Annual Reviews},
}

@article{ash2017elections,
  title={Elections and divisiveness: Theory and evidence},
  author={Ash, Elliott and Morelli, Massimo and Van Weelden, Richard},
  journal={The Journal of Politics},
  volume={79},
  number={4},
  pages={1268--1285},
  year={2017},
  publisher={University of Chicago Press Chicago, IL}
}

@article{saad2021americans,
  title={Americans More Divided on Social Than Economic Issues},
  author={Saad, Lydia},
  journal={Gallup},
  year={2021}
}

\end{singlespace}

\end{document}